\documentclass[aps,prl,showpacs,preprintnumbers,nofootinbib,twocolumn,superscriptaddress]{revtex4-1}

\usepackage{amsmath}
\usepackage{amssymb}
\usepackage{amsfonts}
\usepackage{dsfont}
\usepackage{slashed}
\usepackage{graphicx}
\usepackage{graphics}
\usepackage{epsfig}
\usepackage{bbm,bm}
\usepackage{psfrag}
\usepackage{epstopdf}
\usepackage{hyperref}
\usepackage{color}
\graphicspath{{figures/}}

\usepackage{ulem}
\definecolor{DarkGreen}{rgb}{0.00, 0.60, 0.00}

\newcommand{\tr}{\mathrm{tr}}
\newcommand{\dG}{N^2-1}
\newcommand{\pat}{\partial_t}
\newcommand{\Eqref}[1]{Eq.~\eqref{#1}}

\begin{document}

\title{Asymptotically free scaling solutions in nonabelian Higgs models}

\author{Holger Gies}
\email{holger.gies@uni-jena.de}
\affiliation{\mbox{\it Theoretisch-Physikalisches Institut, Friedrich-Schiller-Universit{\"a}t Jena,}
\mbox{\it D-07743 Jena, Germany}}

\author{Luca Zambelli}
\email{luca.zambelli@uni-jena.de}
\affiliation{\mbox{\it Theoretisch-Physikalisches Institut, Friedrich-Schiller-Universit{\"a}t Jena,}
\mbox{\it D-07743 Jena, Germany}}


\begin{abstract}
We construct asymptotically free renormalization group trajectories
for the generic nonabelian Higgs model in four-dimensional spacetime.
These ultraviolet-complete trajectories become visible by generalizing
the renormalization/boundary conditions in the definition of the
correlation functions of the theory. {Though they are accessible
  in a controlled weak-coupling analysis, these trajectories originate
  from threshold phenomena which are missed in a conventional
  perturbative analysis relying on the deep Euclidean region.}  We
identify a candidate three-parameter family of renormalization group
trajectories interconnecting the asymptotically free ultraviolet
regime with a Higgs phase in the low-energy limit.  We provide
estimates of their low-energy properties in the light of a possible
application to the standard model Higgs sector.  Finally, we find a
two-parameter subclass of asymptotically free Coleman-Weinberg-type
trajectories that do not suffer from a naturalness problem.
\end{abstract}

\maketitle

\section{Introduction}
\label{sec:intro}

While the naturalness problem has been a dominant paradigm for model
building beyond the standard model of particle physics, the triviality
problem of the Higgs sector conceptually appears much more severe as
it inhibits a constructive ultraviolet (UV)-complete definition of the
standard model as an interacting quantum field theory. The triviality
of the standard model Higgs sector is expected to arise from the
fundamental scalar degrees of freedom. For pure scalar theories,
strong evidence for triviality \cite{Wilson:1973jj} -- the fact that
the continuum limit can only be taken for the noninteracting theory --
has been accumulated by lattice simulations in $d=4$
\cite{Luscher:1987ek} and analytic methods \cite{Rosten:2008ts} (see
\cite{Frohlich:1982tw} for a rigorous proof in $d>4$). For nonabelian
Higgs models, Monte-Carlo methods \cite{Lang:1981qg} have found no
indication for continuous phase transitions facilitating a nontrivial
continuum limit.  In practice, triviality arguments have been used to
put upper bounds on the Higgs mass \cite{Maiani:1977cg} long before
its discovery.

Since ATLAS and CMS have found a comparatively light scalar boson
\cite{Aad:2012tfa}, the standard model appears to be in a
``near-critical'' regime \cite{Holthausen:2011aa,Buttazzo:2013uya}
indicating that the Higgs self-interaction is small
near the Planck scale. This
would be natural if all standard-model interactions including the
scalar self-interaction were asymptotically free (AF)
\cite{Gross:1973id}. This is, however, not the case from the standard
viewpoint of perturbative $\beta$ functions.

The construction of AF Yang-Mills-Higgs{(-Yukawa)} systems is in principle
straightforward on the basis of a perturbative analysis
\cite{Gross:1973ju,Chang:1974bv,Fradkin:1975yt,Callaway:1988ya,Giudice:2014tma,Holdom:2014hla}.
In particular, the problematic quartic scalar interaction $\lambda$,
can be marginal-relevant (UV stable) or -irrelevant (UV unstable),
depending on the model and the choice of trajectories.
UV-complete trajectories which emanate from the Gau\ss{}ian fixed
point (FP) can also be built by fixing the unstable marginal-irrelevant
direction.  In RG-improved perturbation theory, this scenario requires
additional fermions as well as eigenvalue conditions
{\cite{Chang:1974bv,Fradkin:1975yt}} to be satisfied
\cite{Salam:1978dk}.  This implies a \textit{reduction of couplings}
\cite{Zimmermann:1984sx}, here effectively removing one parameter, as
$\lambda$ is then purely induced, implying a prediction of the
Higgs-to-$W$-boson mass ratio.  To our knowledge none of such theories
comes sufficiently close to the standard model.  Alternatively, UV
completion in Higgs models can be achieved via asymptotic safety,
which also requires dynamical fermions \cite{Litim:2014uca}.

In this Letter, we consider the construction of AF Yang-Mills-Higgs
systems from a new viewpoint.  Our central idea is that, in order for
suitable AF nonabelian Higgs models to exist, the scalar potential
needs to approach absolute flatness concurrently with the vanishing
gauge coupling $g$. This permits large amplitude fluctuations of
the scalar field controlled by the latter parameter.  We thus suggest to
consider gauge-rescaled scalar field variables $\phi \to g^P \phi$,
with some power $P$, as the relevant measure for amplitudes.  While
$P$ at this point merely seems to be an unphysical rescaling
parameter, we show that it parametrizes RG-scale-dependent boundary
conditions for the effective potential.  These in turn are equivalent
to $g$-dependent renormalization and boundary conditions for the
correlation functions of the theory.  As a consequence, $P$
parametrizes a set of physically distinct RG flows, each one
possessing a Gau\ss ian FP and allowing for AF trajectories.

First signatures of such a trajectory have
been found in a gauged Yukawa model in \cite{Gies:2013pma}. 
In the present work, we explore the general pattern to construct 
UV-complete trajectories for AF nonabelian
Higgs models, including the physically relevant SU(2) model,
for the first time.

\section{A perturbative illustration}
\label{sec:oneloop}

Let us start by recalling the standard perturbative analysis of a
nonabelian Higgs model, as presented, e.g., in ~\cite{Gross:1973ju}.
The one-loop $\beta$-functions derived under the standard assumption
of working in the \textit{deep Euclidean region}, where the RG scale
$k$ is much larger than any other mass scale, read
\begin{equation}
\beta_{g^2}=-b_0 g^4, \quad \beta_{\lambda}=A\lambda^2 +B^\prime\lambda g^2 + C g^4 \ .
\end{equation}
The integrated flow in this simple truncation yields
\begin{equation}\label{eq:integratedflow_oneloop}
\lambda(g^2)=-\frac{g^2}{2A}\left\{B+\sqrt{\Delta}\tanh\!\left[\frac{\sqrt{\Delta}}{2 b_0}\Big(c-\log(g^2)\Big)\right]\right\}
\end{equation}
with $B=B^\prime+b_0$ and $\Delta=B^2-4AC$, and $c$ is an integration
constant.  For the SU(2) model,
\begin{equation*}
b_0=\frac{43}{48\pi^2}\ , \quad A=\frac{3}{4\pi^2}\ , \quad
B^\prime=-\frac{9}{16\pi^2}\ , \quad C=\frac{9}{64\pi^2},
\end{equation*}
such that $\Delta$ is negative and the flow in
\Eqref{eq:integratedflow_oneloop} has a branch cut, the position of
which depends on $c$ and $g^2$. This is the so-called Landau pole,
indicating the unbounded increase of $\lambda$ towards the UV and thus
the failure of perturbation theory. This is considered as reflecting
the triviality problem of the theory which is assumed to persist also
beyond perturbation theory.  If $\Delta$ were positive, $\lambda$
would simply be proportional to $g^2$ itself for sufficiently small
$g^2$ with a $c$-independent proportionality constant. In the limit
$c\to\pm\infty$, two special trajectories would appear, corresponding
to solutions of the FP equation for the ratio
$\zeta=\lambda/g^2$~\cite{Gross:1973ju}
\begin{equation}
\beta_{\zeta}=g^2(A\zeta^2 +B\zeta + C)=0, \quad \zeta=\frac{\lambda}{g^2} \ .
\label{eq:zetapert}
\end{equation}
and they would describe the possible UV asymptotics of all AF
trajectories. Conversely, if at least one of these roots is positive,
then there are AF trajectories in the positive $(g^2,\lambda)$
plane. Nonabelian Higgs models with this property have been
classified, e.g., in \cite{Callaway:1988ya}. The standard SU(2) model
is not of this type.

In order to explore possible loop holes of this conventional
perturbative argument, let us study more general potentials of the form
\begin{equation}
U=\frac{\lambda}{2} \left(\phi^{\dagger}\phi-\frac{v^2}{2}\right)^2 + \frac{\lambda_3}{6k^2} \left(\phi^{\dagger}\phi-\frac{v^2}{2}\right)^3+ \dots .
\label{eq:SSB_truncation_lambda2and3}
\end{equation}
This includes a possible vacuum expectation value $v$ and higher-order
operators such as $\lambda_3$ which can be used to effectively resum
higher loop contributions. A nonperturbative way to study the flow of
general potentials will be used below. Here, we simply study the
contribution of $\lambda_3$ to the flow of $\lambda$. Expressed in
terms of $\zeta$, we find ($\partial_t=d/d\ln k$),
\begin{equation}
\partial_t \zeta=\beta_\zeta=g^2\left(A\zeta^2+B\zeta+C\right)
-\frac{1}{g^2}\left(\frac{\lambda_3}{16\pi^2}+\frac{9\lambda_3}{64\pi^2\zeta}\right)\ .
\label{eq:zeta}
\end{equation}
In contrast to \Eqref{eq:zetapert}, this equation gives rise to a
finite FP value for $\zeta$ if the ratio $\lambda_3/g^4=\chi$ stays
finite and nonvanishing. Note that $\chi$ can have either sign, as
long as the full potential $U$ including higher order terms stays
bounded from below. If realized, this implies that $\lambda$ and
$\lambda_3$ (and possibly all higher $\lambda_{n\geq3}$) are
asymptotically free together with the gauge coupling $g^2$.

Perturbatively, it might seem difficult to stabilize $\lambda_3$ in this
way. However, there is an effect which is missed by the conventional
perturbative analysis: to see this, let us study the flow of the
minimum $v$ of the potential (ignoring wave function renormalizations
for the moment),
\begin{equation}
\partial_t\Big(\frac{v^2}{2k^2}\Big)=-2\Big(\frac{v^2}{2k^2}\Big)
+ \frac{3}{16\pi^2} +\frac{9}{64\pi^2\zeta}.\label{eq:vev}
\end{equation}
For any positive $\zeta$, the ratio $v^2/k^2$ is attracted to a positive UV
fixed point. This implies that $U$ can be attracted towards
a UV fixed point potential in the regime of spontaneous symmetry
breaking (SSB), such that the minimum increases proportional to the RG
scale, $v\sim k$. 

This conclusion has a dramatic consequence: the standard assumption
that the UV behavior of the theory can be exhaustively analyzed in the
deep Euclidean region with $k\gg$ any other scale can be
violated. In order to explore the implications, we have to use a more
powerful formalism that does not rely on the deep Euclidean limit, can
deal with corresponding threshold effects as well as with the RG flow
of full potentials $U$.

Using the functional RG, we show below that the AF scenario visible in
Eqs.~\eqref{eq:zeta}, \eqref{eq:vev} is indeed realized and can be
controlled in a weak-coupling analysis -- though fully accounting for
threshold effects. The scalar couplings run with the gauge coupling to
zero towards the UV, with the ratios of the type $\zeta,\chi$ being
fixed by a boundary condition for the RG flow of the full potential
$U$. In fact, we find a three-parameter family of such RG
trajectories. The flow of the above example with $\lambda_3\sim g^4$
(e.g., $\chi=-2$) and threshold effects included, i.e., $v^2/k^2$ at
its fixed point, is shown below in
Fig.~\ref{fig:streamplot_Phalf_lambda}.

\section{RG flow of the model}
\label{sec:model}

We concentrate on nonabelian Higgs models with a fundamental scalar
$\phi$ as a key building block of the standard model of electroweak
interactions; we consider gauge groups SU($N$), using the standard
model SU(2) for concrete examples. This model includes a Yang-Mills
sector ${\cal L}_\mathrm{YM}=\tr F_{\mu \nu}F^{\mu \nu}/2\ $ with the
field strength $F_{\mu \nu}$ derived from the vector potential
$W_\nu$, and a minimally coupled scalar sector with a scalar potential
that depends on the invariant $\rho:=\phi^{\dagger}\phi$.  In this
work, we analyze the RG flow not only restricted to the set of
perturbatively renormalizable operators, but include a full scalar
potential. Even if the higher operators turn out to be irrelevant and
strongly suppressed along AF trajectories, it is crucial for the UV
construction of these trajectories to go beyond the single-coupling
analysis.  We study the flow of a scale-dependent effective action
\begin{equation}
\Gamma_k=\int Z_{W}{\cal L}_\mathrm{YM}+ Z_{\phi}(D^{\mu}\phi)^{\dagger}(D_{\mu}\phi)+U(\rho),
\end{equation}
where $D_\nu=\partial_\nu-i\bar{g} W_\nu$.  Here, all wave function
renormalizations $Z_{\phi,W}$, the coupling $\bar{g}$, and the
potential $U$ depend on a RG scale $k$.  The RG $\beta$ function(al)s
for these quantities have been computed in \cite{Gies:2013pma}, using
the Wetterich equation~\cite{Wetterich:1992yh,ReviewRG}. This
formulation of the functional RG is useful as it makes no assumptions
about the magnitude of the running masses and couplings, and
incorporates dynamically generated thresholds. The relevance of the
latter for UV completeness has first been studied in
\cite{Gies:2009hq}.

Using the background-field formalism, the running of the renormalized
gauge coupling $g^2=\frac{\bar{g}^2}{Z_W}$ is 
linked to that
of the wave function renormalization \cite{Abbott:1980hw},
\begin{equation}
\pat g^2 = \eta_W\  g^2 ,\quad  \eta_W=-\partial_t\log Z_W, \quad t=\ln k. \label{eq:betagqd}
\end{equation}
The present ansatz for the effective action yields the standard
one-loop running, amended
by threshold effects owing to
gauge bosons and the Higgs scalar acquiring masses in the broken
regime. 
Similarly, the scalar anomalous dimension
$\eta_{\phi}=-\partial_t \log Z_{\phi}$ exhibits
a standard one-loop form including threshold effects
\cite{Gies:2013pma}.

Our search strategy for asymptotic freedom generalizes the
preceding perturbative illustration by looking
for trajectories such that the
$\phi^4$ coupling vanishes as $\lambda\sim g^{4P}$ in the UV, with
arbitrary power $P>0$. The example given above corresponds to
  $P=1/2$.  The nontrivial asymptotic value for $\lambda/g^{4P}$ can
be observed by rescaling the scalar field
\begin{equation}
x=g^{2P}\tilde{\rho}=g^{2P}\frac{ Z_\phi}{k^2}\rho, \quad \rho=\phi^\dagger \phi, \label{eq:gaugerescaling}
\end{equation}
such that $x$ plays the role of a natural
renormalized dimensionless field.
For the full scalar potential, we demand that
higher couplings vanish in the UV
with corresponding or higher powers of $g$.
The dimensionless effective potential
\begin{equation}
f(x)= u(\tilde{\rho})\Big|_{\tilde{\rho}=g^{-2P}x}=k^{-4}U\!\left(\rho\right)\Big|_{\rho=g^{-2P}Z_\phi^{-1}k^2x}\ ,
\end{equation}
should then stay finite and non-vanishing in the far UV 
(the dimensionless quantities $u$ and $\tilde{\rho}$ are often
used in the functional-RG literature).

The flow equation for this rescaled
effective potential reads \cite{Gies:2013pma},
\begin{eqnarray}
\partial_t f &=& \beta_f\equiv -4\, f+(2 + \eta_\phi-P \eta_W)x f' \label{floweq:potential}\\
&&+\frac{1}{16\pi^2} \Big\{3 \sum_{i=1}^{\dG} l_{0\mathrm T}^{(\mathrm{G})4}\left(g^{2(1-P)}\omega_{W,i}^2 (x)\right)\nonumber\\
&&+(2N-1)l_0^{(\mathrm{B})4}\! \left(g^{2P}f' \right)+ l_0^{(\mathrm{B})4}\!\left(g^{2P}(f' + 2 x f'') \right)\!
 \Big\},\nonumber 
\end{eqnarray}
where the scheme-dependent threshold functions $l$ encode the
decoupling of massive modes. Using the linear
regulator~\cite{Litim:2001up}, we have
$l_0^{(\mathrm{B})4}(w)=\frac{1/2}{1+w}\big(1-\frac{\eta_{\phi}}{6}\big)
$ and analogously for $l_{0\mathrm T}^{(\mathrm{G})4}(w)$ upon
replacing $\eta_{\phi}$ by $\eta_{W}$. The gauge-boson mass parameters
$\omega_{W,i}^2 (x)$ arise from the eigenvalues of $(g^{2P}
Z_\phi/k^2) \phi^\dagger\{T^i,T^j\}\phi$, e.g., $\omega_{W,i}^2
(x)=x/2$ for SU(2) for any $i=1,2,3$.

Standard perturbative results are, of course, contained in
\Eqref{floweq:potential}: an expansion to order $\phi^4$ yields the
universal one-loop $\beta_\lambda$ function of \Eqref{eq:zetapert}
upon \textit{(i)} ignoring RG improvement, $\eta_{\phi,W}\to0$ inside
the threshold functions, and \textit{(ii)} taking the deep Euclidean
limit, i.e., ignoring threshold effects
$l_{0}^{(\mathrm{G}/\mathrm{F})4}(w)\to
l_{0}^{(\mathrm{G}/\mathrm{F})4}(0)$ after the expansion in $\phi$.
Similarly, the additional terms $\sim\lambda_3$ in \Eqref{eq:zeta} are
derived by including this operator in the ansatz for the
potential. Projecting onto the flow of the minimum leads to
\Eqref{eq:vev} in the limits \textit{(i)} and \textit{(ii)}. We
emphasize that many of our new results are not fully visible or remain
hidden in this conventional perturbative limit.

\section{Fixed points and scaling solutions}
\label{sec:scaling}

Let us first search for scaling solutions, which correspond to FPs of
the RG flow, representing candidates for asymptotic limits of AF
trajectories. For this, we consider \Eqref{floweq:potential} in the
limit $g\to0$, but keeping $x$ and $f(x)$ finite. The latter
facilitates to consider boundary conditions for the effective
potential, and thus for correlation functions, which are unapparent in
conventional perturbation theory.  Since the scalar loops in the last
line approach irrelevant constants for $g\to 0$, and the anomalous
dimensions also approach zero asymptotically, the flow equation for
$f(x)$ becomes a first-order differential equation. The behavior of
the gauge-boson-loop in the second line, depends on the value of
$P$. For $P\neq 1$, it approaches zero ($P>1$) or an irrelevant
constant ($0<P<1$) and hence can be ignored.  Therefore, for any
regulator and any SU($N$), the FP solutions to the remaining part of
the first line of \Eqref{floweq:potential} satisfying $\partial_tf=0$
read
\begin{equation}
f_*(x)=\xi x^2, \quad P\neq1, \label{eq:FPPneq1}
\end{equation}
for a generic $\xi$ (irrelevant constants in $f(x)$ are
ignored).  For $P=1$, the gauge loop contributes nontrivially to the
effective potential. For SU(2), we find using the linear regulator
\begin{equation}\label{eq:FPpotential}
 f_*(x)= \xi x^2-\left(\frac{3}{16 \pi}\right)^2 \left[2x+
   x^2\log\left(\frac{x}{2+x}\right) \right], \quad P=1,
\end{equation}
with $\xi$ arbitrary. The precise functional form is regulator
dependent, but any regulator yields this Coleman-Weinberg-type
shape. For $\xi\geq0$, the potential is bounded from below and has a
nontrivial minimum ${x_{\text{min}}}_*$.  For $\xi=0$, the minimum is
at infinity. 

The FP potentials of Eqs.~(\ref{eq:FPPneq1},\ref{eq:FPpotential}),
once re-expressed in terms of the original fields $\phi$,  
provide the simplest portrait of a two-parameter family of asymptotically
free solutions. 
Different values of $(\xi,P)$ correspond to different 
flows in coupling space.
This translates into different $g$-dependent
boundary conditions for integrating
the RG equation for $U(\rho)$.
Near the FP, the trajectories differ from
Eqs.~(\ref{eq:FPPneq1},\ref{eq:FPpotential})
by higher powers of the gauge coupling.
The trajectories can systematically be constructed in a weak-coupling expansion
by expanding $\beta_f$ in powers of $g^2$,
and computing the potential $f(x)$
 for which this approximate $\beta$ functional vanishes.
This procedure is justified by the stability analysis given below.

\begin{figure}[!t]
\begin{center}
 \includegraphics[width=0.4\textwidth]{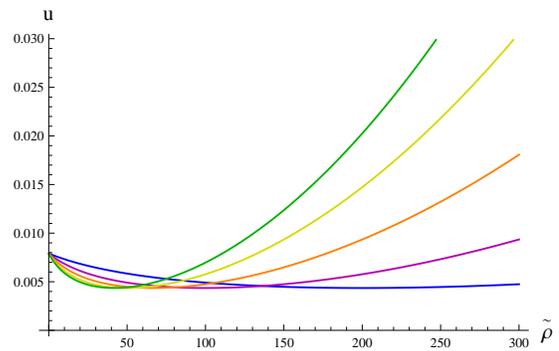}
 \caption{Dimensionless potential $u$ as a function of the
   dimensionless field invariant $\tilde{\rho}$ for the SU($N=2$)
   model with $P=1$ and $\xi\simeq2\times 10^{-4}$ (corresponding to
   $x_{\text{min}*}=2$) for increasing values of $g^2$ from blue (flatter) to
   green (steeper), $g^2 \in \{0.01, 0.02, 0.03, 0.04, 0.048\}$.  }
\label{fig:numerical_marginal_perturbation}
\end{center}
\end{figure}

For $P\in(0,1]$ the next-order approximation includes a linear term in
  the leading power of $g^2$.  The leading power is $g^{2P}$ for
  $P\!\in\!(0,1/2]$, and $g^{2(1-P)}$ for $P\!\in\![1/2,1]$.  The
    corresponding effective potentials are in the SSB regime
\begin{equation*}
 f(x)= \left\{ \begin{array}{ll}
   \xi x^2-\xi \frac{3}{16\pi^2}g^{2P} x & \text{for}\,P\in(0,1/2)\\
   \xi x^2-\frac{3(3+8\xi)}{128\pi^2}g\ x & \text{for}\, P=1/2 \\
   \xi x^2-\frac{9}{128\pi^2}g^{2(1-P)} x & \text{for}\, P\in(1/2,1)
 \end{array}
 \right. .
\end{equation*}
For $P\in(0,1/2)$ or $P=1/2$ the position of the minimum is $g^2$ independent 
($\tilde{\rho}_{\text{min}}=3/32\pi^2$  and 
$\tilde{\rho}_{\text{min}}=3(3+8\xi)/256\pi^2\xi$ respectively), 
whereas for $P\in(1/2,1)$ it is
proportional to $\xi^{-1}g^{2(1-2P)}$ and thus running to infinity in
the UV.  
For $P=1$, we solve the corresponding equation numerically. The
resulting potential $u$ as a function of the unscaled field
$\tilde{\rho}$ is shown in
Fig. \ref{fig:numerical_marginal_perturbation}. Again, the minimum of
$u$ approaches infinity $\sim 1/g^2$ in the UV, and the curvature at
the minimum vanishes like $g^4$.

{While our analysis fully remains in the weak-coupling regime,
  our scaling solutions evade the triviality problem already signalled
  by conventional perturbation theory because of nontrivial threshold
  phenomena: since the scaling potentials have non-trivial minima
  which are finite in dimensionless units or even diverge with
  $g^2\to0$, the threshold effects remain relevant also in the UV. Thus
  the deep Euclidean region which is convenient for a standard
  perturbative analysis is incapable of properly accounting for the
  present scaling solutions.}

{For $P>1$, logarithms slightly complicate the weak-coupling
  expansion. By taking the full gauge loop into account, analytical forms for the scaling solutions can be found which will be given elsewhere \cite{Zambelli:2015}}

\begin{figure}[!t]
\begin{center}
 \includegraphics[width=0.4\textwidth]{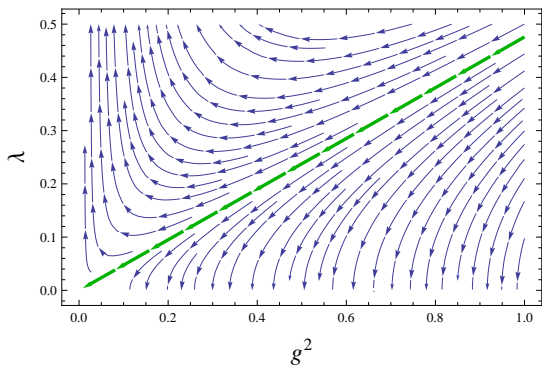}
 \caption{{Phase diagram and UV flow of the SU$(2)$ model in
     terms of $\lambda=u''(\tilde{\rho}_{\text{min}})$ and the gauge
     coupling.  This is obtained by a polynomial truncation of the
     potential $u(\tilde{\rho})$, retaining only the beta functions
     of $\lambda$, $v^2/k^2$ and $g^2$ and their dependence on the higher coupling
     $\chi=u'''(\tilde{\rho}_{\text{min}})/g^4$.  Here we fix $x_{\text{min}}$
     to its FP value and show the
     flow for a constant $\chi=-2$ boundary condition, which is
     consistent with asymptotically-free trajectories for $P=1/2$.
     The colored thick line highlights the root of the finite-$g^2$
     FP equation for the coupling $f''(x_{\text{min}})=\lambda/g^2$.}
 }
\label{fig:streamplot_Phalf_lambda}
\end{center}
\end{figure}

Let us now perform a stability analysis of these trajectories, taking
advantage of their asymptotic description in terms of FPs of the RG
flow of $f(x)$.  For small gauge coupling, perturbations about these
trajectories are translated into deviations from the FP, with components
$\delta g^2=g^2$ and $\delta f(x)=f(x)- f_*(x)$.  Since $\beta_{g^2}$
is proportional to $-g^4$, any eigenperturbation with non-vanishing
gauge coupling must be marginal-relevant. Indeed the $g$-dependent
potential $f$ determined above is by construction a parametrization of
the marginal-relevant eigendirection, since its flow is frozen apart
from the running of $g$.  Conversely, any non-marginal
eigenperturbation must have a vanishing $g^2$ component.  At $g^2=0$,
the eigenvalue problem simplifies to the Gau\ss{}ian one, for which
the eigenperturbations are simple powers, $\delta f \propto x^n$. This
includes a relevant ($n=1$) and a marginal direction ($n=2$).  Beyond
the linear analysis, the $n=2$ direction is actually
marginal-irrelevant, as is familiar from perturbation theory.
{This is visible in the stream-plot of
  Fig.~\ref{fig:streamplot_Phalf_lambda} where the green (thick) line is the projection of a $(P=1/2,\xi\simeq
  0.95)$-asymptotically free trajectory onto the $(g^2,\lambda)$
  plane.}  We emphasize that all our effective potentials are
polynomially bounded, exhibit self-similar eigenperturbations and thus
satisfy standard RG requirements \cite{Morris:1998da}.

To summarize, we have identified new AF trajectories in the nonabelian
Higgs model.  In addition to the standard mass-type relevant
deformation, we have provided the approximate parametrization of one
marginal-relevant eigenperturbation for each pair $(\xi,P)$.
For UV-complete trajectories, the marginal-irrelevant
$\phi^4$-type perturbation is zero.
Therefore, once a specific UV asymptotic behavior is determined by
$(\xi,P)$, only one physical parameter remains apart from an absolute scale.
This is one parameter less than in usual perturbative scenarios. Yet,
we gained the two positive parameters $\xi$ and $P$, labeling
different AF trajectories.

\section{Mass spectrum}

The preceding analysis investigated the UV behavior of the
AF trajectories in the nonabelian Higgs model. In
order to explore the long range mass spectrum, we have to integrate
the flow of the effective potentials towards the IR. If the
trajectories end in a SSB phase, a Fermi scale $k_{\text{F}}$ and
gauge boson and Higgs masses are generated. Trajectories emanating
from a given fixed-point theory specified by $\xi$ and $P$ consist of
the corresponding marginal-relevant eigenperturbation (parametrized by
the gauge coupling $g_\Lambda^2$) possibly superimposed by a finite
component of the relevant direction (the $\delta f\sim x^{n=1}$ Gau\ss{}ian 
perturbation) with some coefficient $c_\Lambda$ at a UV scale
$\Lambda$. Inspired by the standard model hierarchy, we assume
$c_\Lambda$ to be very small, such that the system will spend a long
RG time on top of the marginally relevant trajectory, establishing a
large hierarchy $k_{\text{F}}\ll\Lambda$. At a cross-over (CO) scale
$k_{\mathrm{CO}}$ the relevant component sets in and drives the system
away from the marginal-relevant trajectory. In practice, the initial
conditions $c_\Lambda$, $g_\Lambda^2$ at $\Lambda$ can be traded for
$c_{\mathrm{CO}}$, $g_{\mathrm{CO}}^2$ to be specified at
$k_{\mathrm{CO}}$ (in the standard model, $k_{\text{CO}}\sim
\mathcal{O}(1)$TeV).

For a simple estimate (blind to nonperturbative bound-state effects
\cite{Maas:2012tj}) of the mass spectrum, we initialize the flow at
$k_{\text{CO}}$ with a potential $f_{\mathrm{CO}}$ that is equal to
the 
{analytic parameterization of the marginal perturbation obtained in the previous section,}
plus a relevant component,
$f_{\mathrm{CO}}=f(x; P; \xi; g_{\mathrm{CO}}^2)+ c_{\mathrm{CO}}
x$. We then evolve the full RG flow from $k_{\mathrm{CO}}$ down to
$k_{\mathrm{F}}$.  At the Fermi scale, the gauge coupling $g^2$ as
well as the dimensionful conventionally-renormalized vev $v$ and mass
parameters $m_W^2$, $m_{\mathrm{H}}^2$ in $k_{\mathrm{CO}}$-units,
will depend only on $P$, $\xi$, $c_{\mathrm{CO}}$ and
$g_{\mathrm{CO}}^2$ for sufficiently big $k_{\mathrm{CO}}$ because of
universality. We choose $g_{\mathrm{CO}}^2$ such that
$g^2_{\mathrm{F}}$ acquires a standard-model-like value; since the
gauge running is logarithmically slow, $g_{\mathrm{CO}}^2$ and
$g^2_{\mathrm{F}}$ do not differ significantly.
The parameter $c_{\mathrm{CO}}$ should be chosen sufficiently small in
order to justify that it is ignored above $k_{\text{CO}}$, but also
sufficiently large in order to drive the system rapidly into the SSB
regime; in practice, $c_{\text{CO}}=-0.01$ was used for our
estimates. For the running below $k_{\text{CO}}$, we approximate 
the full effective potential by a 
{standard} polynomial expansion about its
minimum; order-$\phi^8$ polynomials turned out to be sufficient.

{The Higgs-to-gauge boson mass turns out to be an increasing
  function of $\xi$, which is approximately linear, at least for
  small-enough $\xi$, $m_H^2/m_W^2\sim \xi$.  The slope depends on $P$
  and decreases for larger $P$.}  This suggests that any desired
physical value of the mass ratio corresponds to a one-dimensional
section through the $(\xi,P)$ plane, spanning the set of AF
theories. Comparing the IR results at the Fermi scale to the initial
values at $k_{\text{CO}}$, we find that the flow towards the IR
essentially preserves the mass ratio already set by the initial
condition at $k_{\text{CO}}$.  In our scans we observed an almost
$(\xi,P)$-independent ratio $k_\text{CO}/k_\text{F}$ of about one
order of magnitude.

Let us finally explore the physical properties of
Coleman-Weinberg-like trajectories which are defined as those with a
zero relevant component~\cite{Coleman:1973jx}.  {We use an
  order-$\phi^4$ polynomial truncation and integrate the flow by
  keeping fixed the ratio $\chi$ between
  $u'''(\tilde{\rho}_{\text{min}})$ and a suitable power of $g^2$,
  which determines the parameters $(P,\xi)$. To reduce errors, we
  numerically solve the truncated finite-$g^2$ FP equations for $f(x)$
  including subleading corrections to the analytic formulas given
  above.  We observe that these Coleman-Weinberg-like trajectories end
  in the SSB phase in the IR only if the gauge coupling at
  initialization is smaller than a critical $P$-dependent value.  The
  resulting Higgs-to-gauge boson mass parameter ratio is then a
  function of $\xi$.  For instance in the $P=1/2$ case, freeze-out
  occurs when the quartic coupling is still in the FP regime, such
  that the UV relation $\frac{m_{\text{H}}^2}{4m_W^2} =2\xi$ is
  preserved.}

We emphasize that the measured value of the Higgs boson mass can be
understood as essentially driven by top fluctuations
\cite{Gerhold:2009ub,Shaposhnikov:2009pv,Holthausen:2011aa,Gies:2013fua}. The
small Higgs masses (ignoring
  bound-state effects \cite{Maas:2012tj}) in the pure nonabelian
Higgs model along Coleman-Weinberg trajectories thus appear to fit the
requirements of a realistic model. These trajectories may also be
useful to construct a natural large hierarchy in the standard model
via the Higgs portal \cite{Englert:2013gz}; in such a scenario, our
nonabelian Higgs model could play the role of a UV-complete hidden
sector.

In summary, we have discovered a three-parameter family of AF
nonabelian Higgs models. {Our results rely on a controlled
  weak-coupling analysis. Nevertheless, a conventional perturbative
  analysis in the deep Euclidean region is blind to these new
  trajectories as they arise from threshold phenomena which require a
  resummation to become visible in perturbation theory.} If usable in
the context of the full standard model or GUTs, our RG trajectories do
not suffer from $\phi^4$ triviality and thus are candidate building
blocks for a UV-complete quantum field theory. A two-parameter subset
of Coleman-Weinberg-like AF trajectories is even free from the
naturalness problem. We expect these trajectories to be directly
accessible to lattice methods: simulations with bare potentials along
the marginal-relevant eigenperturbations should lie on a line of
constant physics. Still, rather large lattices may be necessary to
resolve the Fermi scale as well as the crossover to the asymptotic
regime.

\acknowledgments

We thank J\"org J\"ackel, Axel Maas, Gian Paolo Vacca and Christof Wetterich for
interesting discussions, and Stefan Rechenberger, Ren\'{e}
Sondenheimer, and Michael Scherer for collaboration on related
projects. We acknowledge support by the DFG under grants
No. GRK1523/2, and Gi 328/5-2 (Heisenberg program).

\end{document}